\documentclass[aps,showpacs,showkeys,twocolumn,amsmath,amssymb]{revtex4-1}

\usepackage{graphicx}
\usepackage{dcolumn}
\usepackage{bm}
\usepackage{mathrsfs}
\usepackage{amsmath}
\usepackage{color}
\usepackage{epsfig}
\usepackage{epstopdf}

\newcommand{\MSPU}{Moscow State Pedagogical University, Moscow 119435, Russia}
\newcommand{\MIPT}{Moscow Institute of Physics and Technology (State University), Dolgoprudny 141701, Russia}
\newcommand{\IMS}{Institute for Physics of Microstructures RAS, Nizhny Novgorod 603950, Russia}
\newcommand{\HSE}{National Research University Higher School of Economics, Moscow 101000, Russia}
\newcommand{\Delft}{Kavli Institute of Nanoscience, Delft University of Technology, Delft 2628 CJ, Netherlands}

\begin{document}
\title{Nonequilibrium interpretation of DC properties of NbN superconducting hot electron bolometers}

\author{M.\,Shcherbatenko}
\affiliation{\MSPU}
\affiliation{\MIPT}

\author{I.\,Tretyakov}
\affiliation{\MSPU}
\affiliation{\MIPT}
\affiliation{\IMS}

\author{Yu.\,Lobanov}
\affiliation{Moscow State Pedagogical University, Moscow 119435, Russia}
\affiliation{Moscow Institute of Physics and Technology (State University), Dolgoprudny 141701, Russia} 

\author{S.\,N.\,Maslennikov}
\affiliation{\MSPU}

\author{N.\,Kaurova}
\affiliation{\MSPU}

\author{M.\,Finkel}
\affiliation{Kavli Institute of Nanoscience, Delft University of Technology, Delft 2628 CJ, Netherlands}

\author{B.\,Voronov}
\affiliation{Moscow State Pedagogical University, Moscow 119435, Russia}

\author{G.\,Goltsman}
\affiliation{\MSPU}
\affiliation{\HSE}

\author{T.\,M.\,Klapwijk}
\email[]{T.M.Klapwijk@tudelft.nl}
\affiliation{\MSPU}
\affiliation{\Delft}

\begin{abstract}
We present a physically consistent interpretation of the dc electrical properties of niobiumnitride (NbN)-based superconducting hot-electron bolometer (HEB-) mixers, using concepts of nonequilibrium superconductivity. Through this we clarify what physical information can be extracted  from the resistive transition and the dc current-voltage characteristics, measured at suitably chosen temperatures, and relevant for device characterization and optimization. We point out that the intrinsic spatial variation of the electronic properties of disordered superconductors, such as NbN, leads to a variation from device to device.
\end{abstract}




\maketitle

Low noise heterodyne-mixers for frequencies beyond 1 THz are based on so-called hot-electron bolometers (HEB). They have been used in the 2009-2013 Herschel Space telescope\cite{graauw2010}, the GREAT instrument on SOPHIA\cite{puetz2012}, and are scheduled to be used in future balloon-borne astronomical experiments\cite{dayton2014}. Unfortunately their physical analysis is poorly developed, which is due to their complexity. First, the absorption of radiation by the superconductor creates a position-dependent non-equilibrium state of the electron-system. Furthermore, in practice the superconductor is coupled to good conducting normal antennas, usually made of gold. Hence, the device is for a certain temperature range essentially a normalmetal-superconductor-normalmetal (NSN) structure, with a low resistance contact between the N and S parts, which means that the proximity-effect plays a role as well as the conversion of normal current to supercurrent. In addition, under operating conditions the superconductor is brought into a superconducting resistive state, which is caused  by a third non-equilibrium phenomenon: the time-dependent changes of the macroscopic superconducting phase. Finally, it has been empirically found that the best devices are obtained by using a very thin film of niobiumnitride (NbN) with a large normal state resistance per square. Such a high degree of disorder competes in a complex way with the tendency to superconduct.\cite{sacepe2008,pratap2013} 
Due to its non-equilibrium position-dependent nature a complete description would require the use of the full machinery of the Keldysh-Usadel theory\cite{Keizer,Vercruyssen}, augmented for the application to strongly disordered superconductors, which is an impossible task. Instead, we present and justify a consistent interpretation and methodology to extract the physically relevant parameters from the dc measurements of practical HEB devices. 
Ultimately the connection to the operation of hot-electron bolometers as mixers should be made, which we do in a separate article.\cite{APL2} Here, we focus only on the physics underlying the dc properties.   

\begin{figure}[t]
\begin{center}
 \includegraphics[width=1\columnwidth]{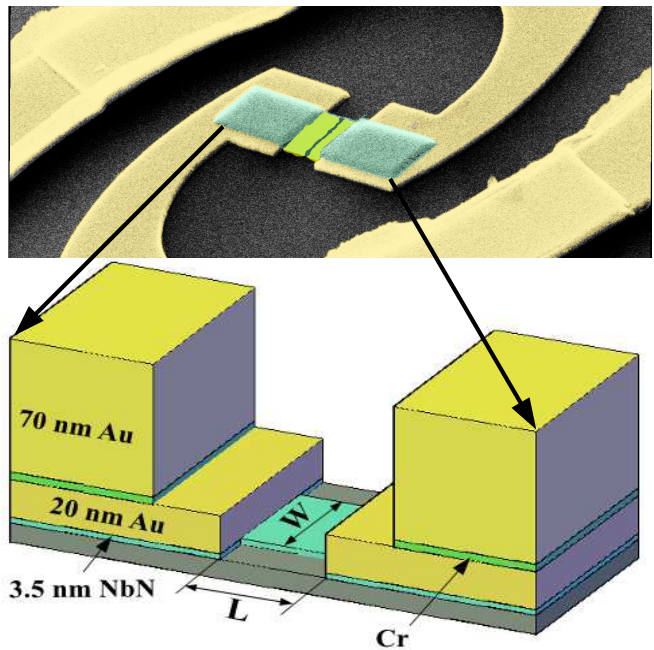}
\end{center}
\caption{\label{fig:3Dview} Upper panel: false color image of a hot-electron bolometer. Lower panel: 
3D view of a typical device.} 
\end{figure}

A typical hot-electron bolometer (Fig.\ref{fig:3Dview}) consists of a centre piece of a thin film of NbN. We study a series of devices made of a 3 to 4 nm thick NbN film shaped to a width $W$ and length $L$. 
In the present experiment we vary intentionally only the width W of the NbN film between the contacts, in order to separate experimentally the properties of the contacts from the properties of the NbN itself. In practice, these contacts provide the boundary conditions for the driven superconducting state in the bare NbN film. As illustrated in the figure the devices consist of 3 different parts. A central, bare NbN film, connected to a NbN-Au bilayer, which on its turn is connected to a NbN-Au-Cr-Au multilayer, which serves as an antenna. During deposition of the NbN the substrate is at a high temperature of $800~ ^0C$. After cooling the substrate to $300~ ^0C$ a thin film of $20\,nm$ thick Au is deposited \emph{in situ}.  This \emph{in situ} process is used to guarantee a good metallic contact between Au and NbN. The Au is subsequently locally removed by ion-etching and wet chemical etching through a window in an electron beam resist, to define the bolometric elements (Fig.\ref{fig:3Dview}).  An additional layer of $70 ~nm$ of Au is deposited to serve as an antenna. This 2nd Au layer is deposited in a separate deposition run, using a few nanometers of Cr for better adhesion. 
We define, $T_{c1}$ as the superconducting transition temperature of the NbN of the active material itself, $T_{c2}$ of the NbN-Au bilayer, and $T_{c3}$ of  the NbN-Au-Cr-Au multilayer. This type of arrangement and fabrication scheme leads to reproducible devices with excellent noise performance when operated as a heterodyne mixer\cite{tretyakov, Baselmans2004}. 
Table \ref{tab:Table1} shows the device parameters.

\begin{table}[t]
  \centering
  \caption{Parameters of the devices made of the same NbN film. Note the increase in the normal resistance per square with decreasing temperature, as well as the variations in $R$ and $T_{c1}$.}
  \label{tab:Table1}
  \begin{tabular}{cccccc}
   
    Device \# & W($\mu m$) & L($\mu m$) & $R_{300} (\Omega/\square)$ & $R_{peak}(\Omega/\square)$ & $T_{c1}(K)$ \\
    \hline
  &  &  &   & &\\
  2 & 0.99 & 0.4 &688 & 1366&9.2\\
  7 & 2.01 & 0.4 & 650 &1276&9.5\\
 8 & 2.01 & 0.4 &685&1328&9.3\\
9 & 2.57 & 0.4  &628&1133&9.6\\
10 &  2.57 & 0.4  &621&1256&9.7\\
11 & 3.12 & 0.4 &601&1102&9.8\\
    \hline
  \end{tabular}
\end{table}

\begin{figure}[t]
\begin{center}
 \includegraphics[width=1\columnwidth]{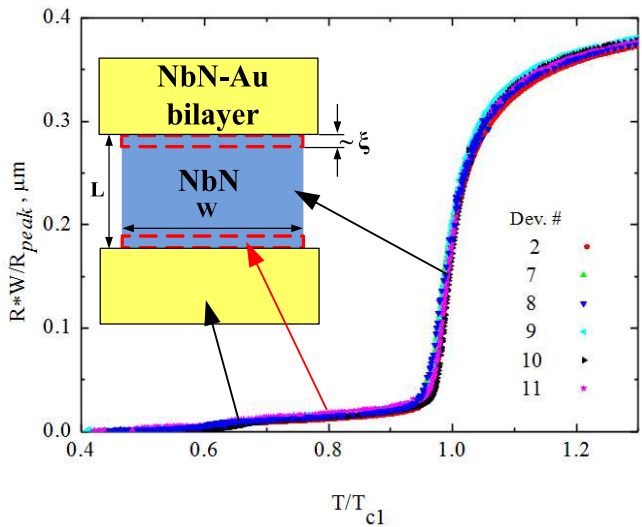}
\end{center}
\caption{\label{fig:RTDevices}Resistance as a function of temperature scaled on the normal state resistance just above $T_{c1}$ and multiplied by the width $W$ for the set of devices listed in Table \ref{tab:Table1}.  Inset: Schematic view of the center of the device. Black arrows indicate which part is determining the critical temperature, $T_{c1}$ for the NbN (blue) and $T_{c2}$ for the bilayer. The dashed red line indicates the superconducting resistive part of the NbN, which causes the plateau-resistance in the $R(T)$ trace indicated with the red arrow.}
\end{figure}

\begin{figure}[t]
\begin{center}
 \includegraphics[width=1\columnwidth]{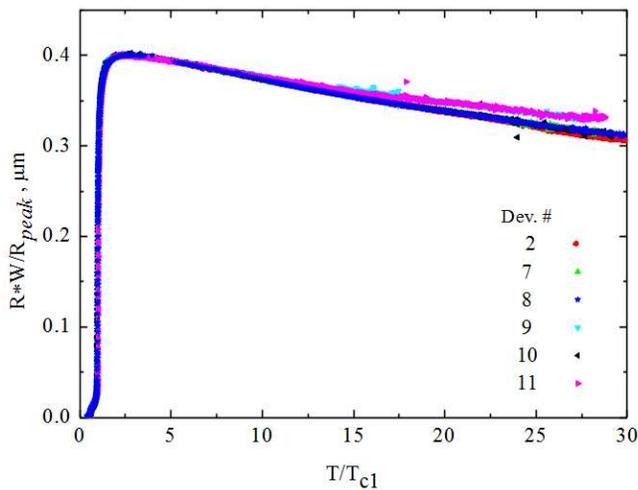}
\end{center}
\caption{\label{fig:SITcurves} Resistive transitions of the devices, all based on a single film of NbN, over a large temperature range. The data are rescaled to the same width and normalized to $R_{peak}$ (Table~\ref{tab:Table1}). } 
\end{figure}

Fig. \ref{fig:RTDevices} shows a set of curves of the resistance as a function of temperature in a narrow temperature range. It reveals the transition curve for uncovered NbN and the end transition of the NbN-Au bilayer. However, it is important to take into account a wider temperature range, which includes the resistance as a function of temperature to room temperature (Fig.\ref{fig:SITcurves}). These curves clearly show a rise in resistance from room temperature to cryogenic temperatures.
The critical temperature $T_{c1}$, taken as the mid-point of the transition, marks the transition temperature of the uncovered NbN of the specific device.
The values for $T_{c1}$ are given in Table \ref{tab:Table1}, and it is clear that they scatter. One also finds a variation in the peak of the resistance just prior to the turn to superconductivity, listed in Table \ref{tab:Table1}, as $R_{peak}$. This variation in $T_{c1}$ and $R_{peak}$ is a 
significant result because these devices are all made of the same film. 
Recent research on strongly disordered superconductors\cite{sacepe2008,pratap2013} suggests that this variation is unavoidable. 
The reason is that the films, approximately $4~nm$ thick,  have a resistance per square of the order of  $1200 ~{\Omega/\square}$, which is equivalent to a resistivity of 480 $\mu\Omega cm$. 
The recent research has made clear that the competition between localization and superconductivity leads in these strongly disordered films, to a spatially fluctuating energy gap, even for atomically uniformly disordered materials\cite{sacepe2008, pratap2013}. 
Hence, superconducting properties need to be determined for each individual device to arrive at a consistent parametrization in the comparision of the devices taking into account the variations in $T_{c1}$ and $R_{peak}$ shown in Table \ref{tab:Table1}. In Fig.\ref{fig:SITcurves} the normal state resistances for all studied devices are shown scaled with the width $W$ and normalized to $R_{peak}$, with temperatures normalized to $T_{c1}$. All curves are on top of each other and in agreement with the lithographically defined length of $0.4~ \mu m$, as is evident from the vertical axis. The value of $R_{peak}$ in Table \ref{tab:Table1} will in the following be taken as the resistivity in the normal state of the superconducting film of the particular device, which will also be a measure of the elastic mean free path and the diffusion constant.
 
The evolution of the resistive superconductivity in the device is apparent from the resistive transition over a much more narrow temperature range around $T_{c1}$ and shown for all devices in Fig. \ref{fig:RTDevices}. The observed resistance is multiplied by the width $W$ and divided by $R_{peak}$ to take out the dependence on the width and the dependence on the resistance in the normal state. We observe clearly two transitions,  a third more gradual transition is in this measurement not clearly visible due to the noise. 
The fact that the normalization on the NbN properties leads to an identical set of curves is a clear indication that we observe systematic behavior for all devices. Since the scaling is based on the width and resistivity of the NbN we must assume that the full stepwise resistive transition of these devices is due to the properties of the NbN and, as often incorrectly assumed, not just only the transition at $T_{c1}$.   
\begin{table}[b]
  \centering
  \caption{Device parameters relevant for the performance.}
  \label{tab:Table2}
  \begin{tabular}{cccccc}
    \toprule
    Device \# &$T_{c1} (K)$&$T_{c2} (K)$&$T_{c3} (K)$&$R_{plateau}(\Omega \mu m )$&$V_c(mV)$ \\
    \hline
&&&&&\\
  2 & 9.2 &5.7  &4.3 &10.9 &1\\
  7 & 9.5 &5.3  &4.7  &12.1 &1\\
 8 & 9.3 &5.6  &4.6&13.1&1.05\\
9 & 9.6 &6.6   &5.5&12.9&1.35\\
10 & 9.7 &6.5   &5.3 &11.6&1.45\\
11 & 9.8 &6.1  &4.8&12.8&1\\
    \hline
  \end{tabular}
\end{table}

With the values of $T_{c1}$ and $T_{c2}$ 
understood, we address the more difficult question of the origin of the observed resistance between $T_{c1}$ and $T_{c2}$ (Table II). 
It has an identical value for all devices if properly scaled on the geometric dimensions of the NbN and the NbN properties. In addition the device is, for that temperature range, an NSN device with the yellow parts in the inset of Fig.\ref{fig:RTDevices} in the normal state and the blue part superconducting. In recent years Boogaard et al\cite{Boogaard} and Vercruyssen et al\cite{Vercruyssen} have 
shown experimentally that in NSN devices an intrinsic contact-resistance arises. In particular in Boogaard et al\cite{Boogaard} it was experimentally proven that the resistance is independent of the length and interpreted as due to the conversion of normal current to supercurrent. For $\Delta \gg k_B T$, far below the critical temperature this conversion-process  occurs with a decay length of about the coherence length. It can be understood as a consequence of Andreev-reflection with diffusely scattered decaying waves at the superconducting side.  We argue below that the plateau resistance in Fig. \ref{fig:RTDevices} is due to this conversion-resistance arising from evanescent states and taking place inside the superconducting NbN over a length of the order of the coherence length, as sketched in the inset of Fig. \ref{fig:RTDevices}.      

We return to the devices shown in Table \ref{tab:Table1} for which we have varied the width $W$ while keeping the length $L$ constant. The width of the NbN-Au bilayer was kept constant, as well as all other parameters. The plateau-resistance occurs between $T_{c1}$ and $T_{c2}$, when the NbN-Au bilayer is normal. The order of magnitude of the resistance is about $10\,\Omega$. 
It could 
be an experimental artefact due to the interface. However, we argue that the resistance is due the uncovered superconducting NbN. 
First, the resistance, which one expects  from the NbN-Au bilayer in the normal state, is in essence equivalent to two parallel resistors of two films with the resistivity of Au and NbN, respectively.  Given the known numbers of their sheet resistances, $2.5\,\Omega/\square$ and $1200\,\Omega/\square$, this resistance is dominated by the 20 nm Au, which quantitatively is much less than the observed value of $10\,\Omega$. The same argument applies to the NbN-Au-Cr-Au multilayer of the antenna. Secondly, as shown in Fig.\ref{fig:RTDevices} all curves collapse onto one curve when scaled with the width $W$ of the NbN, while all other dimensions are kept the same.  Thirdly, if it is due to the same mechanism as reported in Boogaard et al\cite{Boogaard} and Vercruyssen et al\cite{Vercruyssen} it should occur over a length of the order of the coherence length in the NbN. On the vertical axis the data are scaled on the normal state resistance of the device at the peak value just above $T_{c1}$, called $R_{peak}$. Horizontally the temperature is normalized on the $T_{c1}$ of the main resistive transition attributed to NbN.  Fig.~\ref{fig:RTDevices} clearly shows that 
all the curves follow an identical trace, proving that indeed all the properties of the resistive transition are controlled by  the properties of the bare NbN.  From the sheet resistance of  $1200\,\Omega/\square$ we find that the effective length of the two resistive parts of the superconducting NbN is about $8\, nm$. As shown by Boogaard et al\cite{Boogaard} the effective length for 2 interfaces is expected to be equal to $1.92\sqrt{\xi_0 \l}$, with $\xi_0$ the BCS coherence length and $\l$ the elastic mean free path, or about 2 times the low temperature coherence length of a dirty superconductor: $\sqrt{\xi_0 \l}$. Assuming that we can use these results for NbN we find a coherence length of the order of $3\,nm$ to $4\,nm$, in very good agreement with other estimates of the coherence length for NbN. We conclude that for bath temperatures between $T_{c1}$ and $T_{c2}$ the device can be viewed as an NSN device, with the observed plateau-resistance due to the nonequilibrium charge conversion length inside the uncovered NbN. This resistance is realized in the NbN although it is in the superconducting state.  
 Obviously this resistive contribution has a negative effect on the mixing performance of NSN devices because it is only weakly dependent on changes in electron temperature.  
\begin{figure}[t]
\begin{center}
 \includegraphics[width=1\columnwidth]{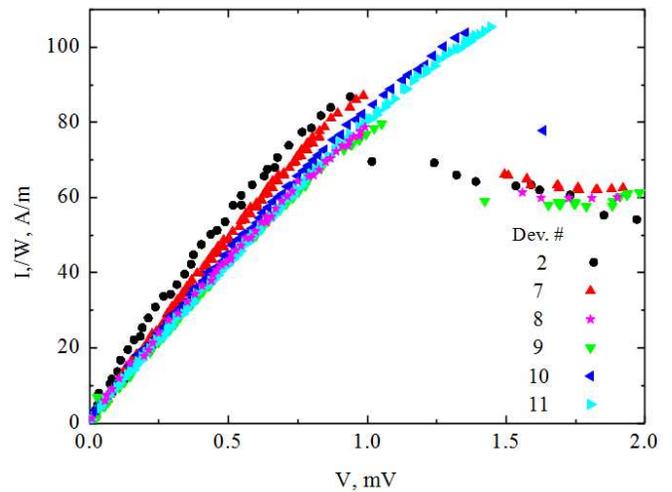}
\end{center}
\caption{\label{fig:ScaledIV}Current-voltage characteristics measured at $T/{T_{c1}}=0.8$ showing that they are nearly linear, indicative of the insensitivity of the conversion resistance to the power delivered to the system. All curves terminate at a specific point in the $I,V$ plane, which is interpreted as a critical voltage arising from the increased energy-mode nonequilibrium.}
\end{figure}

Having identified the origin of the resistance between $T_{c1}$ and $T_{c2}$ it is worth extending the analysis to a study of the current-voltage characteristics in the same temperature range. A typical set, measured at a normalized temperature $T/T_{c1}=0.8$,  is shown in Fig.\ref{fig:ScaledIV}. The $I,V$ curves are almost linear and are terminated at some critical point in the $I,V$ plane, after which the system switches to the normal state. There is some variation from device to device with respect to this critical point, but if properly scaled they are quite similar and the linearity is not trivially expected. The linear behavior indicates that the conversion-resistance at the entry and exit of the superconducting material does not change with increasing bias voltage. Apparently the extra energy which enters the system does not effect the charge-mode of non-equilibrium at the NS interface, which describes the conversion-resistance. However, in addition to the charge mode of nonequilibrium there is also an energy-mode of nonequilibrium, analogous to heating or cooling. Keizer et al\cite{Keizer} have identified a critical voltage at which the superconducting state becomes unstable, for  a voltage approximately equal to $(1/2\sqrt{2})\Delta_0$, with $\Delta_0$ the equilibrium energy gap  of the superconductor. The approach to this point is not observable in the $I,V$ curve because as long as the material stays superconducting, {\emph{i.e.}} carries a supercurrent, it does not contribute to the voltage. However, in the model of Keizer et al\cite{Keizer} and Vercruyssen et al\cite{Vercruyssen} it is assumed that the length of the superconducting wire is short compared to the electron-electron interaction time $\tau_{ee}$, leading to the parameter range $\xi<L<\Lambda_{ee}$. Consequently, a position-dependent non-thermal 2-step distribution function occurs like for normal metal wires studied by Pothier et al.\cite{Pothier1997} For NbN this assumption is not justified because the electron-electron interaction time $\tau_{ee}$ is estimated to be $2.5~psec$ (see Annunziata\cite{Annunziata2010}) or $6.5~psec$ (see Il'in et al\cite{Ilin2000}). For the resistance per square of our samples the diffusion constant $D$ is estimated by rescaling from Semenov et al\cite{Semenov2001} to be $0.2~cm^2/s$, leading to a characteristic length $\Lambda_{ee}$ of $7~nm$ to $12~nm$. Hence, our NbN devices are in a regime where $\xi<\Lambda_{ee}<<L$, which justifies our interpretation of the charge-conversion process.  However, the energy mode of the distribution function has time to become thermal over the length of the superconductor. Hence, it is to be expected that an effective electron temperature\cite{Pothier1997} is given by $T_e(x)=\sqrt{T^2+x(1-x)V^2/L}$.  Here, $T$ is the temperature of the contacts, $V$ the applied voltage, and $L$ the Lorenz number.  The coordinate $x$ runs from 0 to 1 along the superconducting wire. It is to be expected that if $T_e$ is equal to $T_{c1}$ the device will become dissipative at the maximum temperature in the center at $x=L/2$. One expects therefore at $T/T_{c1}=0.8$ that $V_c=1.93~10^{-4} T_{c1}$ with $T_{c1}$ in $K$ and $V_c$ in $V$ in quite good agreement with the data. The fact that $V_c$ is slightly lower can be reconciled by taking into account the reduced heat-diffusion, due to the order parameter profile.\cite{APL2} In addition, some electron-phonon relaxation might also contribute. Assuming that electron-phonon relaxation limits temporal response in mixing experiments\cite{APL2} we infer $\tau_{e-ph}$ of 50 psec, meaning a characteristic length of 35 nm, which has to be taken into account for a full quantitative evaluation. Finally, we emphasize that a critical current in a superconducting nanowire\cite{ant,rom} is a property of a moving Cooper-pair condensate, which is very much different from the critical voltage identified here. The critical pair-breaking current is an equilibrium property, unrelated to the absorption of power. The critical voltage is due to the power fed into the quasi-particle system in a voltage-biased superconductor.                
 
In summary, we have presented an interpretation of the dc properties of hot-electron bolometer mixers, based on nonequilibrium superconductivity. We have identified a temperature-regime in the resistive transition, which can fruitfully be related to the microscopic parameters of NbN.  In addition, we have taken into account that the superconductor is a strongly disordered superconductor with spatially varying superconducting properties. As a consequence a rationale has been given to determine the full set of parameters for each individual device, based on measurements above and below $T_{c1}$, the critical temperature of the NbN but above $T_{c2}$ the critical temperature of the contacts. 
 
\begin{acknowledgments}
We thank J.R.Gao, N.Vercruyssen and S.Ryabchun for stimulating discussions. This work was supported by the Ministry of Education and Science of the Russian Federation, contract N 14.B25.31.0007 of 26 June 2013 and the Russian Science Foundation contract 15-12-10035. TMK also acknowledges the financial support from the European Research Council Advanced grant no. 339306 (METIQUM).  
\end{acknowledgments}



\begin{thebibliography}{9}
\bibitem{graauw2010} Th.~de~Graauw, F.P.~Helmich, T.G.~Phillips, J.~Stutzki, E.~Caux, N.D.~Whyborn, P.~Dieleman, P.R.~Roelfsema, H.~Aarts, R.~Assendorp et al, 
Astronomy \& Astrophysics, {\bf{518}}, L6 (2010)
\bibitem{puetz2012}P.~P\"utz, C.E.~Honingh,K.~Jacobs,M.~Justen, M.~Schultz, and J.~Stutzki, 
Astronomy \& Astrophysics {\bf{542}}, L2 (2012)
\bibitem{dayton2014}D.J.~Hayton, J.L.~Kloosterman, Y.~Ren, T.Y.~Kao, J.R.~Gao, T.M.~Klapwijk, Q.~Hu, C.K.~Walker, J.L.~Reno
Proc. SPIE 9153, Millimeter, Submillimeter, and Far-Infrared Detectors and Instrumentation for Astronomy VII, 91531R (July 23, 2014)
\bibitem{sacepe2008}B.~Sac\'ep\'e, C.~Chapelier, T.I.~Baturina, V.M.~Vinokur, M.R.~Baklanov, and M.~Sanquer
Phys. Rev. Lett. {\bf{101}}, 157006 (2008).
\bibitem{pratap2013}A.~Kamlapure, T.~Das, S.C.~Ganguli, J.B.~Parmar, S.~Bhattacharyya and P.~Raychaudhuri
Scientific Reports {\bf{3}}, 2979 (2013)

\bibitem{Keizer}R.S.~Keizer, M.G.~Flokstra, J.~Aarts, and T.M.~Klapwijk
Phys. Rev. Lett. 96, 147002 (2006)

\bibitem{Vercruyssen}N.~Vercruyssen, T.G.A.~Verhagen, M.G.~Flokstra, J.P.~Pekola, and T.M.~Klapwijk, Phys.Rev. B85, 224503 (2012)
 

\bibitem{APL2}M.~Shcherbatenko, I.~Tretyakov, Yu.~Lobanov, S.N.~Maslennikov, M.~Finkel, A.V.~Semenov, G.~Goltsman, and T.M.~Klapwijk, "Distributed
order parameter model for niobiumnitride hot-electron bolometer mixers"
(unpublished).

\bibitem{tretyakov}I.~Tretyakov, S.~Ryabchun, M.~Finkel, A.~Maslennikova, N.~Kaurova, A.~Lobastova, B.~Voronov, and G.~Gol'tsman, 
Appl. Phys. Lett. 98, 033507 (2011)
\bibitem{Baselmans2004}J.J.A. Baselmans, M. Hajenius, J.R. Gao, T.M. Klapwijk, P.A.J.~de~Korte, B. Voronov, and G. Gol'tsman,
Appl. Phys. Lett. 84, 1958 (2004)
\bibitem{Boogaard}G.R.~Boogaard, A.H.~Verbruggen, W.~Belzig, and. T.M.~Klapwijk
Phys.  Rev. B69, 220503 (2004)

\bibitem{Pothier1997}H.~Pothier, S.~Gu\'eron, N.O.~Birge, D.~Esteve, and M.H.~Devoret, 
Phys. Rev. Lett. 79, 3490 (1997) 

\bibitem{Annunziata2010}A.J.~Annunziata, PhD Thesis, Yale University (2010)
\bibitem{Ilin2000}K.S.Il'in, M.Lindgren, M.Currie, A.D.Semenov, G.N.Gol'tsman and R.Sobolewski
Appl.Phys. Lett. 76, 2752 (2000)
\bibitem{Semenov2001}A.D.~Semenov, G.N.~Goltsman, and A.A.~Korneev,
Physica C 351, 349 (2001)
\bibitem{ant}A.~Anthore, H.~Pothier, and D.~Esteve, 
Phys. Rev. Lett. 90, 127001 (2003)
\bibitem{rom}J.~Romijn, T.M.~Klapwijk, M.J.~Renne, and J.E.~Mooij,
Phys. Rev. B 26, 3648 (1982)
 

\end{thebibliography}
\end{document}